\documentclass[prd,singlecolumn,amsmath,nobibnotes,nofootinbib,superscriptaddress,preprintnumbers,floatfix]{revtex4}

\RequirePackage{epsfig}
\usepackage{graphicx}
\newcommand{\be}{\begin{equation}}
\newcommand{\ee}{\end{equation}}
\newcommand{\ba}{\begin{eqnarray}}
\newcommand{\ea}{\end{eqnarray}}
\newcommand{\bi}{\begin{itemize}}
\newcommand{\ei}{\end{itemize}}

\newcommand{\bfi}{\begin{figure}
\epsfxsize=9cm
\epsffile}
\newcommand{\bfinew}{\begin{figure}
\begin{center}
\includegraphics}
\newcommand{\efi}{\end{figure}}
\newcommand{\efinew}{
\end{center}
\end{figure}}

\graphicspath{ {./figures/} }

\usepackage{epstopdf}
\usepackage{subfigure}
\usepackage[usenames, dvipsnames]{color}
\usepackage{bm} 
\usepackage[linktocpage]{hyperref}
\usepackage{caption}
\usepackage{mathtools}
\def\be{\begin{equation}}
\def\ee{\end{equation}}
\def\ba{\begin{eqnarray}}
\def\ea{\end{eqnarray}}

\def\Tr{\mbox{Tr}}

\newcommand{\wj}[6]{\left(
                           \begin{array}{ccc}
        \! #1\! & #2\!  & #3\!  \\
        \! #4\! & #5\!  & #6\!
                           \end{array}
                   \right)}

\begin{document}

\title{Primordial gravitational wave phenomenology with polarized Sunyaev Zel'dovich tomography}

\author{Anne-Sylvie Deutsch}
\email{asdeutsch@psu.edu}
\affiliation{Institute for Gravitation and the Cosmos and Physics Department, The Pennsylvania State University, University Park, PA 16802, USA}

\author{Emanuela Dimastrogiovanni}
\email{emanuela1573@gmail.com}
\affiliation{Department of Physics/CERCA/Institute for the Science of Origins, Case Western Reserve University, Cleveland, OH 44106, USA}
\affiliation{Perimeter Institute for Theoretical Physics, Waterloo, Ontario N2L 2Y5, Canada}

\author{Matteo Fasiello}
\email{matteo.fasiello@port.ac.uk}
\affiliation{Institute of Cosmology and Gravitation, University of Portsmouth,
Dennis Sciama Building, Portsmouth, PO1 3FX, United Kingdom}
\affiliation{Stanford Institute for Theoretical Physics and Department of Physics, \\Stanford University, Stanford, CA 94306}

\author{Matthew C. Johnson}
\email{mjohnson@perimeterinstitute.ca}
\affiliation{Department of Physics and Astronomy, York University, Toronto, Ontario, M3J 1P3, Canada}
\affiliation{Perimeter Institute for Theoretical Physics, Waterloo, Ontario N2L 2Y5, Canada}

\author{Moritz M{\"u}nchmeyer}
\email{mmunchmeyer@perimeterinstitute.ca}
\affiliation{Perimeter Institute for Theoretical Physics, Waterloo, Ontario N2L 2Y5, Canada}

\preprint{}

\date{\today}

\begin{abstract}
The detection and characterization of primordial gravitational waves through their impact on the polarization anisotropies of the cosmic microwave background (CMB) is a primary science goal of current and future observations of the CMB. An ancillary dataset that will become accessible with the great leaps in sensitivity of CMB experiments is the polarized Sunyaev Zel'dovich (pSZ) effect, small-scale CMB polarization anisotropies induced by scattering from free electrons in the post-reionization Universe. The cross correlation of the pSZ effect with galaxy surveys, a technique known as pSZ tomography, can be used to reconstruct the remote quadrupole field: the CMB quadrupole observed from different locations in the Universe. Primordial gravitational waves leave a distinct imprint on the remote quadrupole field, making pSZ tomography a potential new method to characterize their properties. Building on previous work, we explore the utility of the full set of correlations between the primary CMB and the reconstructed remote quadrupole field to both provide exclusion limits on the amplitude of primordial gravitational waves, as well as to provide constraints on several phenomenological models of the tensor sector: axion gauge field inflation, general models with chiral tensors, and models with modified late-time decay of tensors. We find that relatively futuristic experimental requirements are necessary to provide competitive exclusion limits compared with the primary CMB. However, pSZ tomography can be a powerful probe of the late-time evolution of tensors and, through cross-correlations with the primary CMB, can provide mild improvements on parameter constraints in various models with chiral primordial gravitational waves. 	
\end{abstract}

\maketitle

\section{Introduction}

One of the primary targets of ongoing and future cosmic microwave background (CMB) experiments is the detection of primordial gravitational waves (PGW) through their influence on the polarization anisotropies in the CMB on angular scales of order or larger than a degree \cite{Seljak:1996gy,Kamionkowski:1996zd}. Primordial gravitational waves are a key prediction of inflation, a postulated epoch of accelerated expansion in the early Universe \cite{Guth:1980zm,Lyth:1998xn,Kinney:2003xf,Baumann:2009ds}. The observation of PGW would be an important confirmation of the inflationary paradigm, as well as provide information on the field content of the universe at very high energy scales. However, PGW observations are challenging, requiring high sensitivity, good control of systematics, and precise modelling of foregrounds. This motivates the exploration of alternative ways one might measure the existence and properties of primordial tensors.

Ref.~\cite{2012PhRvD..85l3540A} proposed a new probe of tensors through their effect on the small-scale CMB polarization anisotropies, using the technique of Sunyaev Zel'dovich (pSZ) tomography \cite{Deutsch:2017cja,Deutsch:2017ybc}. The pSZ effect is the polarization anisotropy induced by scattering of CMB photons from electrons in the post-reionization Universe. The imprint of tensors on the pSZ effect arises from the same basic physics as in the primary CMB polarization, namely polarization of scattered CMB photons is induced by a locally observed CMB temperature quadrupole sourced by the time-evolution of tensor modes. The primary CMB polarization traces linear structures at recombination and reionization, while the pSZ effect traces non-linear structures in the late-time Universe. Both depend on the quadrupole field: the CMB temperature quadrupole observed at different points in spacetime. pSZ tomography is the reconstruction of three-dimensional quadrupole field on our past light cone using the correlations between clustering of non-linear structure and the small-scale polarization anisotropies in the CMB.

In this paper, we forecast the potential importance of pSZ tomography in constraining and measuring new physics in the tensor sector. Viewed as a new cosmological observable, the quadrupole field on our past light cone contains new information beyond the primary CMB polarization anisotropies. However, because the quadrupole field has a large correlation length~\cite{Kamionkowski1997,Bunn2006,Portsmouth2004,Hall2014,Deutsch:2017ybc}, any assessment of the importance of this new observable must properly include correlations with the primary CMB. Building on previous work~\cite{2012PhRvD..85l3540A,Deutsch:2017cja,Deutsch:2017ybc}, we perform a series of joint forecasts based on hypothetical CMB experiments and LSS surveys. First, we forecast exclusion limits on the tensor-to-scalar ratio, extending previous work on pSZ tomography~\cite{2012PhRvD..85l3540A} to include information from the primary CMB. We then examine how, in the case of a detection of primordial tensors, pSZ tomography could assist in characterizing new physics in the tensor sector. 

Before proceeding, let us first discuss some possible scenarios for the tensor sector. During inflation, the vacuum fluctuations of the tensor modes are amplified by the expansion, resulting in a stochastic background of gravitational waves. In the simplest, standard, scenario of single-field slow-roll (SFSR) inflation the primordial tensor power spectrum receives its leading contribution from vacuum fluctuations with an equal amplitude for the two polarization states proportional to the Hubble scale, $H$, and a scale-dependence characterized by a small red tilt: 
\begin{equation}
\label{ps}
\mathcal{P}_{h}^{\text{vac}}=\frac{2\,H^2}{\pi^2\,M_{\text{Pl}}^{2}}\left(\frac{k}{k_{*}}\right)^{n_{t}}\,,\quad\quad n_{t}\simeq -2 \epsilon \simeq -r/8\,,
\end{equation}
where $r\equiv\mathcal{P}_{h}/\mathcal{P}_{\zeta}$ is the tensor-to-scalar ratio and $\epsilon\equiv-\dot{H}/H^2$. For a power-law scale-dependence of the tensor power spectrum, from the largest CMB scales all the way to interferometer scales, current bounds require $-0.62< n_{t}<0.53$ for $r_{0.01}<0.080$ \cite{Akrami:2018odb}.

Vacuum fluctuations, however, only account for the homogeneous solution to the full equation of motion for primordial tensor modes
\begin{eqnarray}\label{eqh}
\ddot{h}_{ij}(\tau,\vec{k})+3\,H\,\dot{h}_{ij}(\tau,\vec{k})+k^2 h_{ij}(\tau,\vec{k})=\frac{2}{M_{P}^{2}}\,\mathcal{S}^{TT}_{ij}(\tau,\vec{k})\,,
\end{eqnarray}
where $h_{ij}$ is the traceless and transverse part of the metric and $\mathcal{S}^{TT}_{ij}$ is the source term. Whilst in SFSR the inhomogeneous solution to Eq.~(\ref{eqh}) can be safely neglected, if additional fields are present during inflation these will contribute to $\mathcal{S}^{TT}_{ij}$ and the resulting inhomogeneous solution should be taken into consideration as it can, in many scenarios, dominate the homogeneous one. The resulting power spectrum can be rather different than that of Eq.~(\ref{ps}), including the possibility of turning blue ($n_{t}>0$) and displaying ``bump"-like features at specific scales.

The source term arises due to extra content in the inflationary Lagrangian, from scalars to higher-spin particles. In particular, gauge fields exhibit an interesting phenomenology. Many pages of what one might call the Encyclopedia Inflationaris have been written about vector gauge fields, in most cases a $U(1)$ field identifying a preferred direction and thus providing a mechanism for generating (controlled) anisotropic observables (see e.g. \cite{Bartolo:2012sd}). Vector modes tend to decay during inflation unless their coupling to the inflaton is promoted to be non-minimal so that the inflaton field can ``lift" the extra degrees of freedom. This tendency to decay early on  can be turned into a virtue whenever the inflaton potential at hand is too steep to sustain a sufficiently long expansion and/or deliver the measured scalar spectral index. In such cases, the direct coupling to gauge fields serves as a dissipation channel, flattening the slope of the inflaton effective potential. It is in this latter role that gauge fields have often been employed within axion inflation models.

The appeal of axion models relies in no small part on their ability to solve the $\eta$-problem for inflation: an (approximate) shift symmetry protects the inflaton mass from large quantum corrections. These properties were already all in place in the first of such proposals, \textit{natural inflation} \cite{Freese:1990rb}. However, if one aims at ensuring a sufficiently mild slope for the inflationary effective potential without resorting to a trans-Planckian field-excursion, coupling to gauge fields has proven extremely effective (see e.g. \cite{Kim:2004rp} for alternatives).  It is in this context that we will consider here the so-called axion-gauge field (A-G) models.

The Lagrangian describing the generic class of A-G theories typically (see e.g. \cite{Anber:2009ua,Barnaby:2010vf,Sorbo:2011rz,Barnaby:2011vw,Cook:2011hg,Adshead:2012kp,Martinec:2012bv,Barnaby:2012xt,Noorbala:2012fh,Dimastrogiovanni:2012st,Dimastrogiovanni:2012ew,Adshead:2013nka,Mukohyama:2014gba,Ferreira:2014zia,Obata:2014loa,Namba:2015gja,Obata:2016tmo,Peloso:2016gqs,Dimastrogiovanni:2016fuu,Garcia-Bellido:2016dkw,Fujita:2017jwq,Caldwell:2017chz,Agrawal:2017awz,Agrawal:2018mrg,Lozanov:2018kpk,Dimastrogiovanni:2018xnn,McDonough:2018xzh,Papageorgiou:2018rfx,Domcke:2018rvv}) contains the coupling $\chi F\tilde{F}$ ($\chi$ being the axion and $F$ the field strength for the gauge modes), for the simple reason that this is allowed by the symmetries of the Lagrangian (and may therefore appear after quantum corrections are taken into account).
The presence of this interaction term is responsible for one remarkable phenomenological implication for PGW: the rolling axion enhances \textit{one} polarization of the gauge-field fluctuations which, in turn, sources the corresponding polarization of gravitational waves. The resulting PGW have a chiral tensor power spectrum with a scale-dependence that may exhibit a bumpy feature at a given scale (see e.g. \cite{Barnaby:2012xt,Mukohyama:2014gba,Namba:2015gja,Peloso:2016gqs,Dimastrogiovanni:2016fuu})\footnote{We do not commit to a specific scale here because one of the models we shall consider (see Sec.~\ref{cnf}) has enough freedom in the parameter space to accommodate the bump-like feature in different regimes (e.g. CMB scales, or scales of interest for interferometers).}. Non-trivial chirality and scale-dependence are then two key properties that can be used to distinguish vacuum-sourced PGW from those that receive the leading contribution from gauge fields. One should also mention non-Gaussianity: it has been shown that some A-G models generate potentially observable tensor as well as mixed scalar-tensor bispectra (see e.g. \cite{Agrawal:2017awz,Agrawal:2018mrg,Dimastrogiovanni:2018xnn}). 

In another class of models, tensors are sourced by spectator scalar fields with small sound speed. These spectator fields contribute to the source term in Eq.~(\ref{eqh}) at quadratic level and their effect on the primordial signal can be as large as that from vacuum fluctuations \cite{Biagetti:2013kwa,Biagetti:2014asa,Fujita:2014oba}. If the sound speed is time-dependent, the spectral index $n_{t}$ can be blue~\footnote{There is at least one other interesting route to a positive tensor tilt. The standard inflationary dynamics breaks time-diffeomorphism invariance as a result of the time-evolving background of the inflaton field. Space-diffs are preserved to guarantee homogeneity and isotropy of the background. Intriguingly, this specific symmetry breaking pattern is not strictly necessary for a successful inflationary mechanism. Starting with solid inflation \cite{Endlich:2012pz}, various viable models breaking space-diffs have been  proposed \cite{Cannone:2014uqa,Cannone:2015rra,Ricciardone:2016lym}. In many of these set-ups a mass term is generated for the graviton leading to the possibility of a blue PGW spectrum (see e.g. \cite{Bartolo:2016ami} for forecasts on the parameter space of these models for LISA).} in light of the contribution from the parameter $s\equiv\dot{c}_{s}/H c_s$ and support PGW growing towards smaller scales \cite{Khoury:2008wj,Bartolo:2016ami}. It is worth mentioning that, from the model building point of view, scalars with a reduced sound speed can result from integrating out kinetically coupled heavy fields such as in the mechanism of \cite{Tolley:2009fg} (see also \cite{Achucarro:2012yr,Burgess:2012dz} and references therein).

Cosmological observables are sensitive to both the primordial spectrum of tensors as well as their late time evolution. Within $\Lambda$CDM, the evolution of tensors is entirely fixed by the expansion history. However, there exist a number of modifications of general relativity, invoked to address dark energy, which can alter the late-time evolution of tensors. A vanilla example is quintessence, where the decay of tensors is different than for a pure cosmological constant. Of course, for any observationally viable theory of quintessence, the modification to decay is small. A more exotic example arises in theories of massive bigravity~\cite{Hassan:2011zd} (see also~\cite{deRham:2010kj}), where the growth of tensors at late times can be strongly affected by new degrees of freedom~\cite{Cusin:2014psa,Johnson:2015tfa,Fasiello:2015csa}. 

In this paper, we focus on a handful of phenomenological models which encompass a variety of the scenarios described above, assessing the impact of pSZ tomography on parameter constraints. First, we consider the exclusion bounds on the tensor-to-scalar ratio $r$ in the absence of tensors. We then consider models with a power law spectrum as in Eq.~(\ref{ps}), including the possibility of non-zero chirality, and forecast constraints on $r$, $n_t$, and the chirality parameter $\Delta_c$ for a variety of fiducial choices of the model parameters. This result complements previous work constraining chiral tensors in the primary CMB~\cite{Saito:2007kt,Gluscevic:2010vv,Gerbino:2016mqb}. Next we consider a three-parameter family of models describing chiral tensors sourced by gauge fields. This result complements forecasted constraints on such models using the primary CMB~\cite{Thorne:2017jft,Hiramatsu:2018nfa}. Finally, we consider a phenomenological model of modified late-time growth of tensors.

The paper is organized as follows. In Sec.~\ref{sec:signal}, we review the contribution from tensors to the primary CMB temperature and polarization anisotropies and the remote quadrupole field. In Sec.~\ref{sec:forecastsetup} we outline the framework and assumptions for our forecasts. In Sec.~\ref{sec:rlimits} we present exclusion limits on $r$ and in Sec.~\ref{cnf} we present our forecasted parameter constraints on two classes of models. We conclude in Sec.~\ref{conclusions}.

\section{Tensor contributions to the remote quadrupole field and CMB anisotropies}\label{sec:signal}

In this section, we review the contributions from tensors to the quadrupole field and to the primary CMB temperature and polarization anisotropies. We also review the reconstruction noise expected on the quadrupole field given a LSS survey and CMB experiment. We refer readers to Refs.~\cite{2012PhRvD..85l3540A,Deutsch:2017cja,Deutsch:2017ybc} for further details.

\subsection{Tensor modes}

We decompose the tensor modes into $+$ and $\times$ polarizations as:
\begin{equation}
	g_{ij} = a^2 \left(\begin{array}{ccc} 1+h_+ & h_\times & 0 \\ h_\times & 1-h_+ & 0 \\ 0 & 0 & 1 \end{array}\right).
\end{equation}
In Fourier space, the time-evolution of the tensor modes is encoded in the tensor transfer function $D^T$:
\begin{equation}
h_{(+,\times)} (k,\chi) = D^T (k,\chi) h_{i,(+,\times)}(k)\,.
\end{equation}
Below, we approximate the tensor transfer function by 
\begin{equation}
 D^T (k,\chi) = 3 \frac{j_1 (k \chi)}{k\chi},
\end{equation}
valid in a purely matter-dominated Universe. Within $\Lambda$CDM, tensors experience additional decay at late times due to the presence of a cosmological constant. We neglect this effect for simplicity in most of what follows, but asses the sensitivity of pSZ tomography to altered late-time decay in Sec.~\ref{sec:alteredgrowth} using a one-parameter phenomenological model.

The correlators for $h_{i,(+,\times)}(k)$ are defined by
\begin{eqnarray}
\langle h_{i,+}(k) h_{i,+}^*(k') \rangle &=& (2 \pi)^3  \delta^{(3)} (k-k') \frac{1}{2} \left[ P_L + P_R \right]  = \frac{1}{2} P_h\,, \\
\langle h_{i,\times}(k) h_{i,\times}^*(k') \rangle &=& (2 \pi)^3  \delta^{(3)} (k-k') \frac{1}{2} \left[ P_L + P_R \right] = \frac{1}{2} P_h\,, \\
\langle h_{i,+}(k) h_{i,\times}^*(k') \rangle &=& (2 \pi)^3  \delta^{(3)} (k-k') \frac{i}{2} \left[ P_L - P_R \right] = i \frac{\Delta_c}{2} P_h\,.
\end{eqnarray}
Here, $P_L$ and $P_R$ corresponds to the power in left- and right-handed circularly polarized gravitational waves, and $P_h$ is the total power in tensor modes. We have allowed for chirality, described by the parameter $\Delta_c$, through the definitions 
\begin{equation}
P_L = \frac{1}{2} \left[ 1+ \Delta_c \right] P_{h}, \ \ \ P_R = \frac{1}{2} \left[ 1- \Delta_c \right] P_{h}\,.
\end{equation}
The chirality parameter takes the values $-1 \leq \Delta_c \leq 1$. We define the dimensionless total tensor power spectrum by:
\begin{equation}
P_{h} = \frac{2 \pi^2}{k^3} \mathcal{P}_h\,.
\end{equation}
{For a power-law scaling, as is the case for single-field inflation, one has
\begin{equation}\label{eq:rnt}
\mathcal{P}_h = \frac{A_t}{2} \left( \frac{k}{k_0}\right)^{n_t}\,.
\end{equation}}
We choose the pivot scale $k_0 = 0.05 \ {\rm Mpc}^{-1}$. The factor of 2 arises from our choice of normalization for the gravitational wave polarization vectors.

\subsection{Temperature and polarization anisotropies}

\subsubsection{Primary CMB temperature}

The contribution from tensors to the spherical harmonic coefficients of the primary CMB temperature is 
\begin{equation}\label{eq:almT}
a_{\ell m}^T = - \int \frac{d^3 k}{(2 \pi)^3} \ \mathcal{I}^T_{\ell}(k,\chi=0) \left\{ h_{i,+}(k) \left[{}_{2}Y^{*}_{\ell m}({\bf \widehat{k}}) + {}_{-2}Y^{*}_{\ell m}({\bf \widehat{k}}) \right]  \right.  + i \left. h_{i,\times}(k) \left[{}_{2}Y^{*}_{\ell m}({\bf \widehat{k}}) - {}_{-2}Y^{*}_{\ell m}({\bf \widehat{k}}) \right] \right\}\,,
\end{equation}
where $\mathcal{I}^T_{\ell}(k,\chi=0)$ is defined as
\begin{equation}\label{eq:Iintegral}
\mathcal{I}^T_{\ell}(k,\chi(a_e)) = \pi \sqrt{\ell (\ell^2-1)(\ell+2)} \int_{a_e}^{a_\text{dec}} da \frac{d D^T_{}(k,a)}{da} \ \frac{j_\ell (k\Delta\chi(a))}{[k\Delta\chi(a)]^2}\,.
\end{equation}
Here, $\chi(a_e)$ is the radial comoving distance to a scatterer and $\Delta\chi(a) = -\int_{a_e}^{a} da' [H(a')a'^2]^{-1}$. For the primary CMB temperature, $\chi(a_e) = 0$.

\subsubsection{Remote quadrupole field}

The remote quadrupole field is the CMB temperature quadrupole as viewed from different locations in spacetime, defined as:
\begin{equation}
q_{\pm} ({\bf\hat{n}} ,\chi) \equiv \sum_{m=-2}^{2} a_{2m}^T ({\bf \hat{n}}, \chi) \left._{\pm 2}Y_{2 m}\right. ({\bf \hat{n}})\,.
\end{equation}
Here, $a_{2m}^T ({\bf \hat{n}}, \chi)$ are the moments of the temperature quadrupole observed at position ${\bf r} = \chi {\bf \hat{n}}$. Decomposing into spin-2 harmonics $a_{\ell m}^{\pm q} (\chi)$, and using the usual definition of curl-free $E$ and curl $B$ modes, we define the $E$-mode and $B$-mode remote quadrupole fields, which are given by:
\begin{align}
	a_{\ell m}^{qE}(\chi) = & \int \frac{d^3 k}{(2\pi)^3}\ 5 i^\ell B_{\ell}(k,\chi) \mathcal{I}^T_{2}(k,\chi) \left\{ h_{i,+}(k) \left[{}_{2}Y^{*}_{\ell m}({\bf \widehat{k}}) + {}_{-2}Y^{*}_{\ell m}({\bf \widehat{k}}) \right]  \right.  + i \left. h_{i,\times}(k) \left[{}_{2}Y^{*}_{\ell m}({\bf \widehat{k}}) - {}_{-2}Y^{*}_{\ell m}({\bf \widehat{k}}) \right] \right\}, \label{eq:talmE}\\
	a_{\ell m}^{qB}(\chi) = & - \int \frac{d^3 k}{(2\pi)^3}\ 5 i^\ell A_{\ell}(k,\chi)  \mathcal{I}^T_{2}(k,\chi) \left\{h_{i,+}(k) \left[{}_{2}Y^{*}_{\ell m}({\bf \widehat{k}}) - {}_{-2}Y^{*}_{\ell m}({\bf \widehat{k}}) \right]  \right.  + i \left. h_{i,\times}(k) \left[{}_{2}Y^{*}_{\ell m}({\bf \widehat{k}}) + {}_{-2}Y^{*}_{\ell m}({\bf \widehat{k}}) \right] \right\}. \label{eq:talmB}
\end{align}
Here, $A_{\ell}(k,\chi)$ and $B_{\ell}(k,\chi)$ are:
\begin{eqnarray} \label{eq:Al_def}
	A_{\ell}(k,\chi)&\equiv&\frac{1}{2}\left(\frac{2 j_{\ell}(k\chi)}{k\chi}+\frac{1}{k}\frac{d}{d\chi}j_{\ell}(k\chi)\right)  \,,\\ 
	B_{\ell}(k,\chi)&\equiv& -\frac{1}{4k^2}\frac{d^2 j_{\ell}(k\chi)}{d\chi^2}-\frac{1}{k^2\chi}\frac{d j_{\ell}(k\chi)}{d\chi}+j_{\ell}(k\chi)\left(\frac{1}{4}-\frac{1}{2(k\chi)^2}\right)  \,.  \label{eq:Bl_def}
\end{eqnarray}
These have the limiting values of $A_{\ell}(k,\chi\rightarrow 0)\rightarrow 0$ and $B_{\ell}(k,\chi\rightarrow 0)\rightarrow -1/5$. The integral $\mathcal{I}^T_{2}(k,\chi)$ is defined above in Eq.~(\ref{eq:Iintegral}). Comparing Eq.~(\ref{eq:talmE}) with Eq.~(\ref{eq:almT}), note that $a_{2 m}^{qE}(\chi \rightarrow 0) \rightarrow a_{2 m}^T$. Note further that $a_{\ell m}^{qB}(\chi \rightarrow 0) \rightarrow 0$. This is as it should be: the remote quadrupole field evaluated at our position is simply the locally observed temperature quadrupole. 

In practice, one cannot perfectly reconstruct the full redshift dependence of the remote quadrupole fields. One strategy, employed in previous work, is to reconstruct an averaged quadrupole field in a set of redshift bins. We define:
\begin{equation}
a_{\ell m \alpha}^{qE} \equiv \int d\chi \ W(\chi, \chi_\alpha) \  a_{\ell m}^{qE}(\chi), \ \ \ \ a_{\ell m \alpha}^{qB} \equiv \int d\chi \ W(\chi, \chi_\alpha) \  a_{\ell m}^{qB}(\chi)\,,
\end{equation}
where $W(\chi, \chi_\alpha)$ is a set of unit-norm top-hat functions in comoving distance labeled by the index $\alpha = 1, 2, \ldots N_{\rm bins}$. Below, we choose a configuration with 12 bins of equal size in comoving distance between $0.1\leq z \leq 6$. This choice of coarse-graining of the quadrupole field retains most of the relevant information for the parameter constraints presented below.  

\subsubsection{Primary CMB polarization}

The tensor contribution to CMB polarization is:
\begin{equation}
(Q \pm iU)({\bf \hat{n}}) = \frac{\sqrt{6}}{10}  \int d\chi \ g(\chi) \ q_{\pm} ({\bf\hat{n}} ,\chi).
\end{equation}
where $g(\chi)$ is the visibility function. We calculate the visibility function in the vicinity of recombination using the recfast++ code~\cite{2010MNRAS.403..439R,Chluba:2010ca,doi:10.1111/j.1365-2966.2009.15957.x,doi:10.1111/j.1365-2966.2010.16940.x,Seager:1999bc}; we further assume instantaneous reionization consistent with an optical depth of $\tau= 0.06$. Expanding this expression in spin-2 harmonics and defining $E$ and $B$ modes, we can use the result above for the remote quadrupole field to obtain:
\begin{eqnarray}\label{eq:primary_E}
a_{\ell m}^{E} &=& \frac{\sqrt{6}}{10}  \int d\chi \ g(\chi) \ a_{\ell m}^{q,E} (\chi) \ , \\
a_{\ell m}^{B} &=& \frac{\sqrt{6}}{10}  \int d\chi \ g(\chi) \ a_{\ell m}^{q,B} (\chi)\,,
\end{eqnarray}
where $a_{\ell m}^{q,E} (\chi)$ is defined in Eq.~(\ref{eq:talmE}) and $a_{\ell m}^{q,B} (\chi)$ is defined in Eq.~(\ref{eq:talmB}). 

\subsection{Power spectra}

We can define a set of transfer functions $\Delta_{\ell}^X (\chi)$ to compute the various angular power spectra for the E-mode remote quadrupole field $qE$ and B-mode remote quadrupole field $qB$ (both functions of $\chi$) as well as the primary CMB $T$, $E$, and $B$ modes. In the absence of chirality ($\Delta_c = 0$), the non-zero correlations are:
\begin{eqnarray}
C_{\ell, \alpha \alpha'}^{XX}  &=& \int d \ln k \ \Delta_{\ell \alpha}^X (k) \Delta_{\ell \alpha'}^X (k) \ \mathcal{P}_h, \ \ \ X = \{qE, qB\} \\
C_{\ell, \alpha}^{qE X} &=&  \int d \ln k \ \Delta_{\ell \alpha}^{qE} (k) \Delta_{\ell}^X (k) \ \mathcal{P}_h, \ \ \ X=\{ T,E\} \\
C_{\ell,\alpha}^{qB B} &=&  \int d \ln k \ \Delta_{\ell \alpha}^{qB} (k) \Delta_{\ell}^B (k) \ \mathcal{P}_h, \\
C_{\ell}^{X Y} &=&  \int d \ln k \ \Delta_{\ell}^{X} (k) \Delta_{\ell}^Y (k) \ \mathcal{P}_h, \ \ \ X,Y = \{ T,E\} \\
\end{eqnarray}
In the presence of chirality ($\Delta_c > 0$), there are the additional correlations:
\begin{eqnarray}
C_{\ell, \alpha \alpha'}^{qEqB} &=&  \Delta_c \int d \ln k \ \Delta_{\ell \alpha}^{qE} (k) \Delta_{\ell \alpha'}^{qB} (k) \ \mathcal{P}_h, \\
C_{\ell, \alpha}^{qE B} &=&  \Delta_c \int d \ln k \ \Delta_{\ell \alpha}^{qE} (k) \Delta_{\ell}^B (k) \ \mathcal{P}_h, \\
C_{\ell \alpha}^{qB X} &=&  \Delta_c \int d \ln k \ \Delta_{\ell \alpha}^{qB} (k) \Delta_{\ell}^X (k) \ \mathcal{P}_h, \ \ \ X=\{ T,E\} \\
C_{\ell}^{XB} &=&  \Delta_c \int d \ln k \ \Delta_{\ell}^{X} (k) \Delta_{\ell}^{B} (k) \ \mathcal{P}_h, \ \ \ X=\{ T,E\} \\
\end{eqnarray}
The transfer functions appearing in these expressions are given by:
\begin{eqnarray}
\Delta_{\ell \alpha}^{qE} (k) &=& \frac{5}{\sqrt{2 \pi}} \int d\chi' \ W(\chi',\chi_\alpha) \ B_{\ell}(k,\chi') \mathcal{I}^T_{2}(k,\chi')\,, \\
\Delta_{\ell \alpha}^{qE} (k) &=& -\frac{5}{\sqrt{2 \pi}} \int d\chi' \ W(\chi',\chi_\alpha) \ A_{\ell}(k,\chi') \mathcal{I}^T_{2}(k,\chi') \,,\\
\Delta_\ell^{E} (k) &=& - \frac{\sqrt{6}}{10} \int d\chi' \ g(\chi') \ \frac{5}{\sqrt{2 \pi}} B_{\ell}(k,\chi') \mathcal{I}^T_{2}(k,\chi') \,,\\
\Delta_\ell^{B} (k) &=& \frac{\sqrt{6}}{10} \int d\chi' \ g(\chi') \ \frac{5}{\sqrt{2 \pi}} A_{\ell}(k,\chi') \mathcal{I}^T_{2}(k,\chi') \,,\\
\Delta_\ell^{T} (k) &=& - \frac{1}{\sqrt{2 \pi}} \mathcal{I}^T_{\ell}(k,0)\,.
\end{eqnarray}

\begin{figure}[t!]
  \includegraphics[width=3.in]{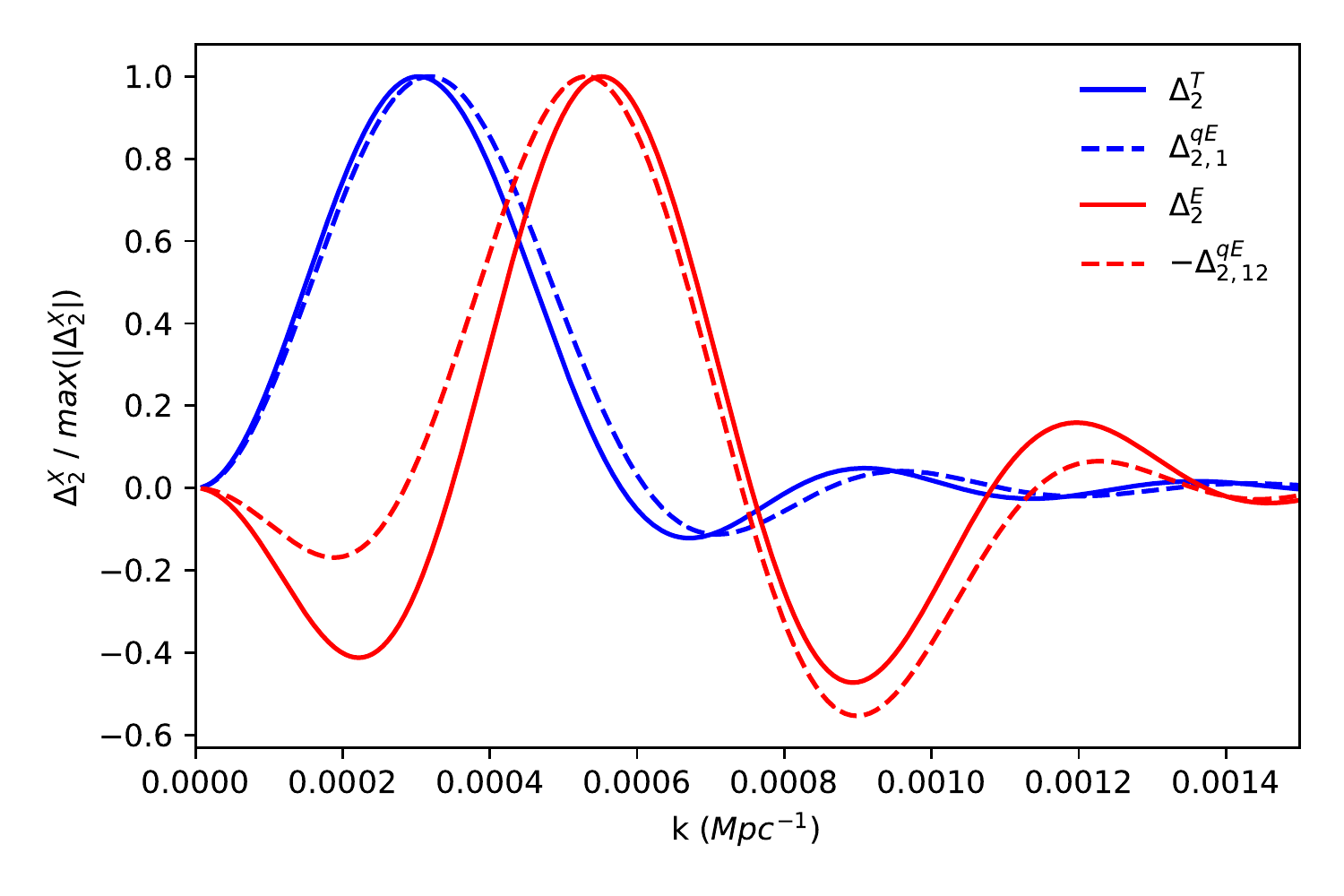}
  \includegraphics[width=3.in]{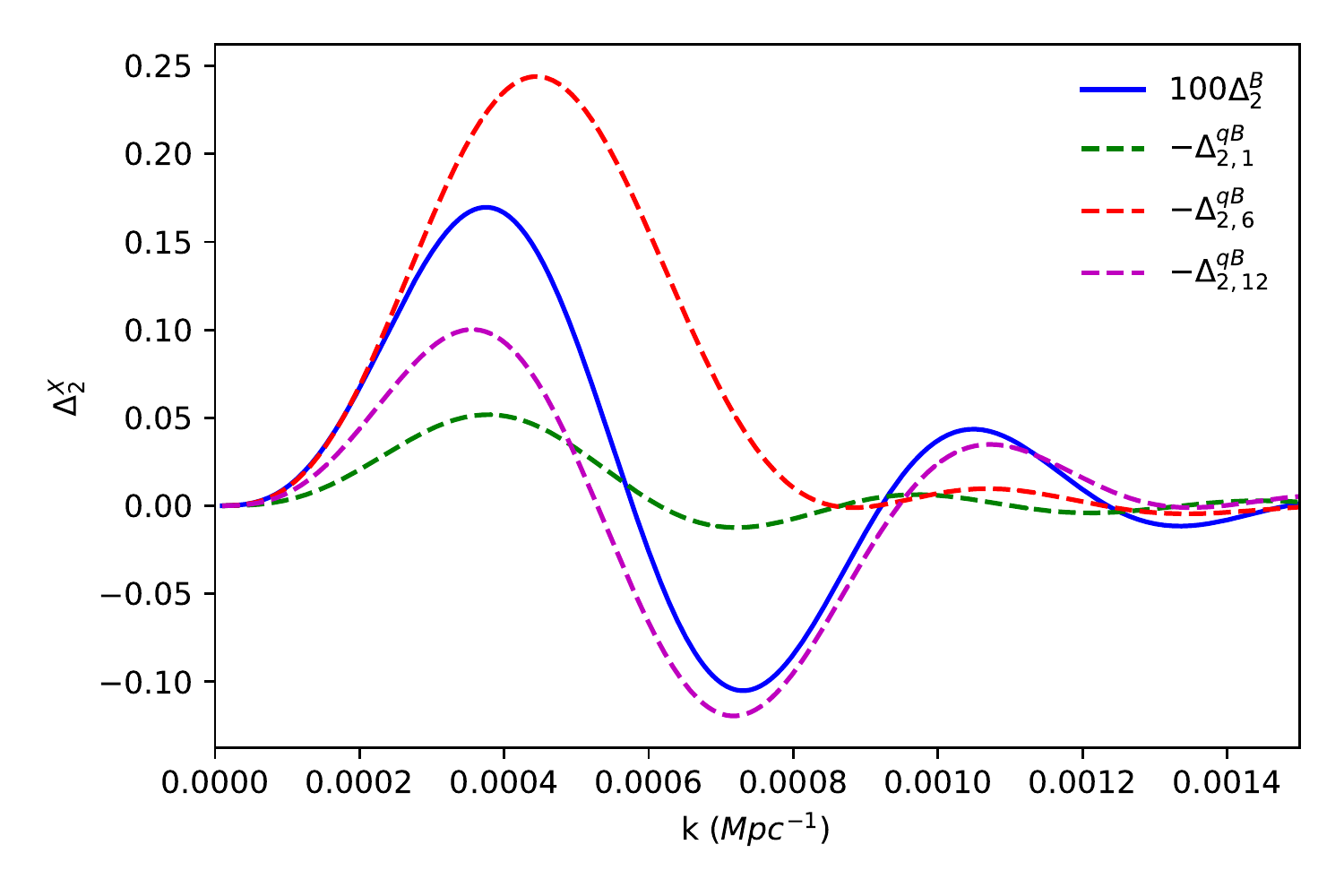}
	\caption{Transfer functions defining the remote quadrupole $qE$ (left) and $qB$ fields (right) in different redshift bins as compared to the primary CMB transfer functions. Note that the transfer functions in the left panel are normalized to their maximum value, while the transfer functions in the right panel are not. On the left, we see a strong overlap in shape between $qE$ in the lowest redshift bin and $T$ as well as between $qE$ in the highest redshift bin and $E$. On the right, we demonstrate that the transfer function has a maximum amplitude at intermediate redshifts, as well as the overlap in shape between $qB$ in the highest redshift bin and $B$.}
	\label{fig:transferTEqE}
\end{figure}

To gain some understanding of the importance of various correlations, we compare in Fig.~\ref{fig:transferTEqE} the transfer functions for qE and the primary CMB T/E (left panel) as well as qB and the primary CMB B (right panel). First, as described above, there is a strong correlation between the $\ell=2$ moments of CMB T and qE in the lowest redshift bin. There is also a correlation between CMB E and qE in the highest redshift bin. This could be anticipated from Eq.~(\ref{eq:primary_E}), since the visibility  
function weights the integral over the remote quadrupole towards higher redshifts. Likewise, there is a significant overlap between the transfer functions for qB at high redshift and CMB B modes. Note also that the amplitude of the $qB$ transfer function is largest at intermediate redshifts. This is a consequence of the competition between vanishing $	A_{\ell}$ as $\chi \rightarrow 0$ (see Eq.~(\ref{eq:Al_def})) and the decrease in $\mathcal{I}^T_{2}(k,\chi)$ at large $\chi$.

\subsection{CMB and reconstruction noise}\label{sec:noise}

In the forecasts presented below, we consider the effect of instrumental noise associated with the CMB experiment as well as the reconstruction noise on the remote quadrupole field. We assume Gaussian CMB instrumental noise and a Gaussian beam
\begin{equation}\label{eq:cmbnoise}
N_\ell^{X} = (\Delta T)^2 \exp\left[ \frac{\ell (\ell+1) \theta_{\rm FWHM}^2}{8\ln 2} \right]\,,
\end{equation}
where $\theta_{\rm FWHM}$ is the beam width and $\Delta T$ the amplitude of the instrumental noise, with the noise for $T,E,B$ assumed to be identical. We assume $\theta_{\rm FWHM}=1 \ {\rm arcmin}$ and varying $\Delta T$ for our CMB experiment. The reconstruction noise on the remote quadrupole field is given by~\cite{Deutsch:2017ybc}:
\ba
\label{eq:estimator1noiseq}
\frac{1}{N^{qE \alpha}_{\ell}} &=& \frac{1}{(2\ell+1)} \sum_{\ell_1\ell_2}
\frac{\Gamma^{\rm pSZ}_{\ell \ell_1\ell_2 \alpha} \ \Gamma^{\rm pSZ}_{\ell \ell_1\ell_2 \alpha}}{\Big( |\alpha_{\ell,\ell_1,\ell_2}|^2 C^{EE}_{\ell_1} + |\gamma_{\ell,\ell_1,\ell_2}|^2 C^{BB}_{\ell_1} \Big) C^{\delta_g \delta_g}_{\alpha \ell_2} }\,,\\
\frac{1}{N^{qB \alpha}_{\ell}} &=& \frac{1}{(2\ell+1)} \sum_{\ell_1\ell_2}
\frac{\Gamma^{\rm pSZ}_{\ell \ell_1\ell_2 \alpha} \ \Gamma^{\rm pSZ}_{\ell \ell_1\ell_2 \alpha}}{\Big( |\gamma_{\ell,\ell_1,\ell_2}|^2 C^{EE}_{\ell_1} + |\alpha_{\ell,\ell_1,\ell_2}|^2 C^{BB}_{\ell_1} \Big) C^{\delta_g \delta_g}_{\alpha \ell_2} }\,, 
\label{eq:estimator2noiseq}
\ea
where
\ba
F_{\ell,\ell_1,\ell_2} &=& \sqrt{\frac{(2\ell+1)(2\ell_1+1)(2\ell_2+1)}{4\pi}} \wj{\ell}{\ell_1}{\ell_2}{2}{-2}{0}\\
\alpha_{\ell,\ell_1,\ell_2} &=& \frac{1}{2} (1 + (-1)^{\ell+\ell_1+\ell_2}) \\
\gamma_{\ell,\ell_1,\ell_2} &=& \frac{1}{2i} (1 - (-1)^{\ell+\ell_1+\ell_2}) \\
\Gamma^{\rm pSZ}_{\ell \ell_1 \ell_2 \alpha} &=& - \frac{\sqrt{6}}{10} F_{\ell,\ell_1,\ell_2} \ C^{\Delta\tau \delta_g}_{\alpha,\ell_2}.
\ea
$C^{EE}_{\ell}, \ C^{BB}_{\ell}$ are the lensed polarization power spectra computed using CAMB assuming cosmological parameters from Planck 2015 (CMB only)~\cite{Ade:2015xua}, and including instrumental noise defined by Eq.~(\ref{eq:cmbnoise}). The reconstruction noise in Eq.~(\ref{eq:estimator1noiseq},\ref{eq:estimator2noiseq}) is computed as in Ref.~\cite{Deutsch:2017ybc}. Briefly, $C^{\delta_g \delta_g}_{\alpha \ell}$ is the binned galaxy-galaxy power spectrum, which is computed in the Limber approximation to be
\begin{equation}
C^{\delta_g \delta_g}_{\alpha \ell} =  \int \frac{dk}{\ell+\frac{1}{2}} W(\chi, \chi^\alpha)^2 b_{g}(z[\chi])^2 P_{mm}(k,\chi) \Big|_{\chi \rightarrow (\ell+1/2)/k} + \frac{1}{N_{g, \alpha}}
\end{equation}
where we have assumed that the galaxy power spectrum is related by a redshift-dependent linear bias $b_{g}(z)^2$ to the non-linear matter power spectrum $P_{mm}$ (computed using CAMB~\cite{Lewis:1999bs,Howlett:2012mh}). The second term accounts for shot noise in the galaxy survey, where $N_{g, \alpha}$ is the number of galaxies per square radian in the bin labeled by $\alpha$. We consider a galaxy sample consistent with a large photometric redshift survey performed by the Large Synoptic Survey Telescope (LSST)~\cite{LSSTScienceCollaboration2009}, with a number density given by
\begin{equation}
\label{eq:lsstnr}
N_{g, \alpha} = \int_{\chi_\alpha^{\rm min}}^{\chi_\alpha^{\rm max}}  d\chi \ n(z[\chi]), \ \ \ \ n(z) =  n_{\rm gal} \ \frac{1}{2 z_0} \left(\frac{z}{z_0}\right)^2 \exp(-z/z_0), \ \ \   
\end{equation}
with $z_0 = 0.3$ and $n_{\textrm{gal}}=40 \ {\rm arcmin}^{-2}$. We model the galaxy bias by
\begin{equation}
b_{g}(z) = \frac{0.95}{D(z)}
\end{equation}
where $D(z)$ is the matter growth function, computed using CAMB. We assume that the distribution of electrons traces that of dark matter on all scales, and in the Limber approximation we obtain
\begin{equation}
C^{\tau \delta_g}_{\alpha \ell} = \sigma_T  \int \frac{dk}{\ell+\frac{1}{2}} \left(a(\chi) \bar{n}_e(\chi) \right) W(\chi, \chi^\alpha) b_g(z[\chi]) P_{mm}(k,\chi) \Big|_{\chi \rightarrow (\ell+1/2)/k}.
\end{equation}
for the optical depth-galaxy cross power. We assume full-sky experiments and evaluate the reconstruction noise in the range $100 \leq \ell \leq 10^4$. This treatment neglects a number of potentially important physical effects, such as baryonic feedback on the distribution of electrons, partial sky overlap between the CMB and LSS experiment, and non-linear galaxy bias. These features will be explored in future work.

Beyond the instrumental and reconstruction noise, there are contributions to the various auto- and cross-power spectra from scalar modes. In addition to the usual scalar contributions to the CMB temperature, the scalar remote quadrupole contributes to $qE$ and $E$, and lensing contributes to $B$. However, $qB$ provides a 'clean' measurement of tensors since it does not receive any contribution from scalar modes. We refer the reader to Refs.~\cite{2012PhRvD..85l3540A,Deutsch:2017cja,Deutsch:2017ybc} for a detailed discussion of the scalar contributions to the remote quadrupole; we employ these results in our forecast below.

\begin{figure}
  \includegraphics[width=3.in]{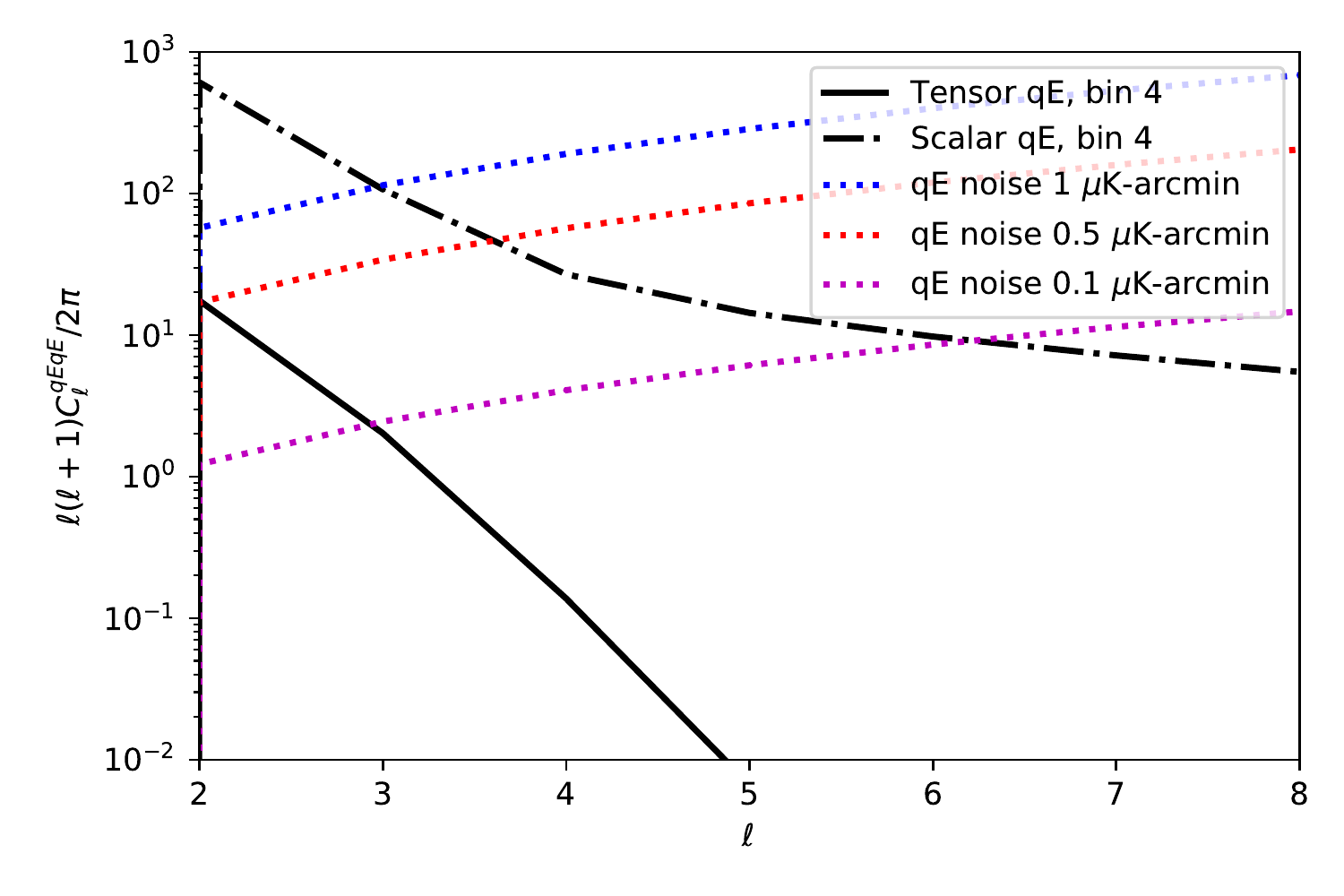}
  \includegraphics[width=3.in]{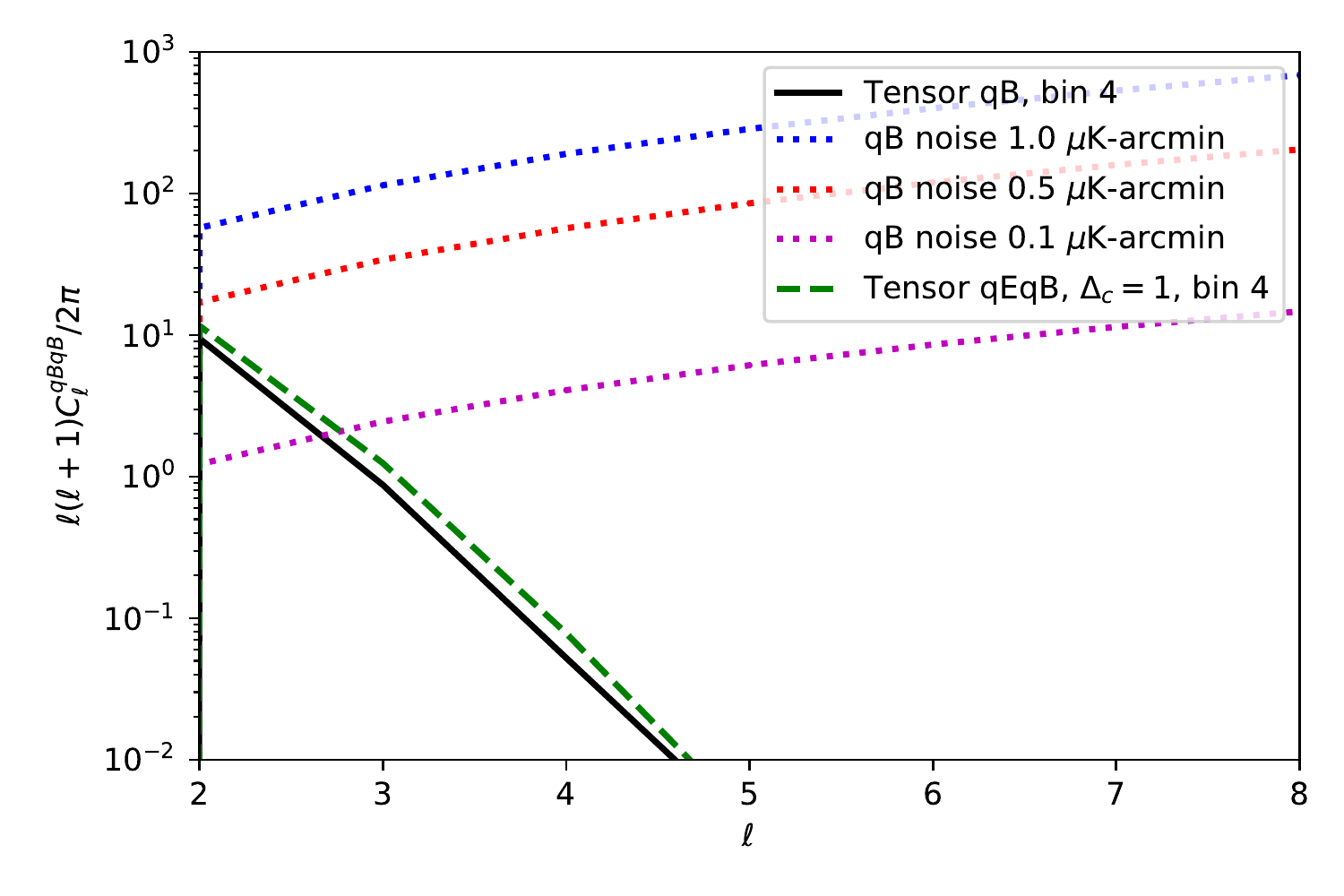}
	\caption{The remote quadrupole $qE$ (left) and $qB$ (right) power spectra (black solid) for an intermediate redshift bin assuming $r=0.05$. Overplotted is the reconstruction noise assuming data on the full-sky and an LSST-like galaxy survey for varying levels of instrumental noise (dotted lines). On the left, we also show the contributions to $qE$ from scalar modes (black dot-dashed). On the right, we show the $qEqB$ cross power (green dashed) for a maximally chiral model, $\Delta_c=1$. }
	\label{fig:recnoise}
\end{figure}

In Fig.~\ref{fig:recnoise}, we compare the tensor contribution to $C_\ell^{qEqE}$ (left panel) and $C_\ell^{qBqB}$ (right panel) for $r = 0.05$ to the scalar contribution (dot-dashed line) and reconstruction noise (dotted lines) assuming an LSST-like galaxy survey and a full-sky CMB experiment with 1 arcmin beam and $\{1, 0.1, 0.01 \} \ \mu$K-arcmin instrumental noise. The scalar contribution to $C_\ell^{qEqE}$ dominates the tensor contribution over all angular scales for this value of $r$, making this a major contaminant for the detection of tensors with $qE$. It can be seen that a sensitivity around $0.5 \ \mu$K-arcmin is necessary to detect the $\ell=2$ modes of $qB$ at an SNR of order one. To capture modes with $\ell > 2$ at significant SNR, it is necessary to drop the reconstruction noise by orders of magnitude. To determine what is necessary to do so, we can examine the asymptotic noise-limited behavior of Eq.~(\ref{eq:estimator1noiseq}). Consider the limit where $C^{EE}_{\ell}$ and $C^{BB}_{\ell}$ are dominated by instrumental noise, neglecting the beam so that $C^{EE}_{\ell} \sim C^{BB}_{\ell} \sim \Delta T^2$, and where $C^{\delta_g \delta_g}_{\alpha \ell}$ is dominated by shot noise so that $C^{\delta_g \delta_g}_{\alpha \ell} \sim 1/N_{g, \alpha}$. In the noise-limited regime, we have $N^{(qE,qB) \alpha}_{\ell} \propto \Delta T^2/N_{g, \alpha}$. Therefore, either increasing the sensitivity of the CMB experiment or increasing the size of the redshift catalog will lower the reconstruction noise. However, because the tensor contribution to the remote quadrupole power is a steeply falling function of $\ell$, large improvements in sensitivity and survey size are necessary to add even a few measurable modes. 

Let us pause here to illustrate an additional point. In Fig.~\ref{fig:recnoise}, we plot the $qEqB$ cross-power (green dashed) for a maximally chiral model, $\Delta_c=1$, with $r=0.05$. This falls between the $qEqE$ and $qBqB$ power, as one would have naively expected. This is in contrast to the $EB$ cross-power in the primary CMB, which is suppressed~\cite{Masui:2017fzw} by a factor of $\sim 10^{-2}$ relative to the $EE$ power due to projection effects. Therefore, we see that pSZ tomography exhibits the increased sensitivity to chirality of tensors associated with a three-dimensional probe~\cite{Masui:2017fzw}.

\section{Forecast setup}\label{sec:forecastsetup}

In this section we describe the framework and assumptions that go into our forecasts. As datasets, we include hypothetical measurements of the primary CMB $T, E, B$ as well as the remote quadrupole $qE$ and $qB$ fields reconstructed using pSZ tomography, as described above. We consider the full set of correlation functions between CMB $T, E, B$ and remote quadrupole $qE$ and $qB$ in the multipole range $2 \leq \ell_{\rm low} \leq 10$, including both scalar and tensor contributions. At higher-$\ell$, the amplitude of $qE$ and $qB$ is undetectably small, and so we consider the primary CMB $T, E, B$ alone (including both scalar and tensor contributions) in the multipole range $10 < \ell_{\rm high} \leq 1000$. We neglect the possibility of de-lensing the primary CMB $B$-modes. For the range of $r$ currently allowed, the lensing $B$-modes comfortably dominate the tensor contribution on the upper end of the multipole region we consider, justifying our cutoff at $\ell = 1000$ for this forecast. Including the possibility of delensing would affect our results, both for the primary CMB and remote quadrupole constraints, and we defer an exploration of this possibility to future work. For the CMB experiment, we fix the beam size to $1$ arcmin and vary the instrumental noise. We neglect foregrounds and systematics and assume data on the full sky.

Our forecasts are performed in the context of the Fisher matrix formalism. The Fisher matrix is
\begin{equation}
F_{ij} =  \sum_\ell \frac{2\ell+1}{2} \Tr \left[ (\partial_i {\bf C}_\ell) {\bf C}_\ell^{-1} (\partial_j {\bf C}_\ell) {\bf C}_\ell^{-1} \right]
\end{equation}
where $i$ and $j$ are the free parameters in the tensor model, described below. The covariance matrix ${\bf C}_\ell$ includes the tensor and scalar signals as well as instrumental and reconstruction noise. For the low-$\ell$ range of multipoles, with both the primary CMB and remote quadrupole fields, we have
\begin{equation}
{\bf C}_\ell = {\bf C}_\ell^{{\rm low}-\ell}  + {\bf N}_\ell^{{\rm low}-\ell}
\end{equation}
with
\begin{equation}
{\bf C}_\ell^{{\rm low}-\ell} = \begin{pmatrix} 
C_\ell^{TT} & C_\ell^{TE} & C_\ell^{TB} & C_{\ell,\alpha_1}^{TqE} & \ldots & C_{\ell,\alpha_N}^{TqE} & C_{\ell \alpha_1}^{TqB} & \ldots & C_{\ell \alpha_N}^{TqB} \\
C_\ell^{TE} & C_\ell^{EE} & C_\ell^{EB} & C_{\ell,\alpha_1}^{EqE} & \ldots & C_{\ell,\alpha_N}^{EqE} & C_{\ell \alpha_1}^{EqB} & \ldots & C_{\ell \alpha_N}^{EqB} \\
C_\ell^{TB} & C_\ell^{EB} & C_\ell^{BB} & C_{\ell,\alpha_1}^{BqE} & \ldots & C_{\ell,\alpha_N}^{BqE} & C_{\ell \alpha_1}^{BqB} & \ldots & C_{\ell \alpha_N}^{BqB} \\
C_{\ell, \alpha_1}^{TqE} & C_{\ell, \alpha_1}^{EqE} & C_{\ell, \alpha_1}^{BqE} & C_{\ell, \alpha_1 \alpha_1}^{qEqE} &  \ldots & C_{\ell, \alpha_1 \alpha_N}^{qEqE} & C_{\ell, \alpha_1 \alpha_1}^{qEqB} & \ldots & C_{\ell, \alpha_1 \alpha_N}^{qEqB} \\
\ldots & \ldots & \ldots & \ldots & \ldots & \ldots & \ldots & \ldots & \ldots \\
C_{\ell, \alpha_N}^{TqE} & C_{\ell, \alpha_N}^{EqE} & C_{\ell, \alpha_N}^{BqE} & C_{\ell, \alpha_N \alpha_1}^{qEqE} & \ldots & C_{\ell, \alpha_N \alpha_N}^{qEqE} & C_{\ell, \alpha_N \alpha_1}^{qEqB} & \ldots & C_{\ell, \alpha_N \alpha_N}^{qEqB}\\
C_{\ell, \alpha_1}^{TqB} & C_{\ell, \alpha_1}^{EqB} & C_{\ell, \alpha_1}^{BqB} & C_{\ell, \alpha_1 \alpha_1}^{qEqB} &  \ldots & C_{\ell, \alpha_1 \alpha_N}^{qEqB} & C_{\ell, \alpha_1 \alpha_1}^{qBqB} & \ldots & C_{\ell, \alpha_1 \alpha_N}^{qBqB} \\
\ldots & \ldots & \ldots & \ldots & \ldots & \ldots & \ldots & \ldots & \ldots \\
C_{\ell, \alpha_N}^{TqB} & C_{\ell, \alpha_N}^{EqB} & C_{\ell, \alpha_N}^{BqB} & C_{\ell, \alpha_N \alpha_1}^{qEqB} & \ldots & C_{\ell, \alpha_N \alpha_N}^{qEqB} & C_{\ell, \alpha_N \alpha_1}^{qBqB} & \ldots & C_{\ell, \alpha_N \alpha_N}^{qBqB}\\
\end{pmatrix}
\end{equation}
where the various spectra include relevant scalar and tensor contributions. The noise term is
\begin{equation}
{\bf N}_\ell^{{\rm low}-\ell} = \begin{pmatrix} 
N_\ell^{TT} & 0 & 0 & 0 & \ldots & 0 & 0 & \ldots & 0 \\
0 & N_\ell^{EE} & 0 & 0 & \ldots & 0 & 0 & \ldots & 0 \\
0 &0 & N_\ell^{BB} &0 & \ldots & 0 & 0 & \ldots & 0 \\
0 & 0 & 0 & N_{\ell, \alpha_1}^{qEqE} &  \ldots & 0 & 0 & \ldots & 0 \\
\ldots & \ldots & \ldots & \ldots & \ldots & \ldots & \ldots & \ldots & \ldots \\
0 & 0 & 0 & 0 & \ldots & N_{\ell, \alpha_N}^{qEqE} & 0 & \ldots & 0 \\
0 & 0 & 0 & 0 &  \ldots & 0 & N_{\ell, \alpha_1}^{qBqB} & \ldots & 0 \\
\ldots & \ldots & \ldots & \ldots & \ldots & \ldots & \ldots & \ldots & \ldots \\
0 & 0 & 0 & 0 & \ldots & 0 & 0 & \ldots & N_{\ell, \alpha_N}^{qBqB}\\
\end{pmatrix}
\end{equation}
For the high-$\ell$ range of multipoles, with only the primary CMB, we have
\begin{equation}
{\bf C}_\ell = {\bf C}_\ell^{{\rm high}-\ell}  + {\bf N}_\ell^{{\rm high}-\ell}
\end{equation}
with
\begin{equation}
{\bf C}_\ell^{{\rm high}-\ell} = \begin{pmatrix} 
C_\ell^{TT} & C_\ell^{TE} & C_\ell^{TB} \\
C_\ell^{TE} & C_\ell^{EE} & C_\ell^{EB} \\
C_\ell^{TB} & C_\ell^{EB} & C_\ell^{BB} \\
\end{pmatrix}
\end{equation}
where again both scalar and tensor contributions are included in each of the spectra. The noise term is:
\begin{equation}
{\bf N}_\ell^{{\rm high}-\ell} = \begin{pmatrix} 
N_\ell^{TT} & 0 & 0 \\
0 & N_\ell^{EE} & 0 \\
0 & 0 & N_\ell^{BB} \\
\end{pmatrix}
\end{equation}
The marginalized constraints on the free parameters of the various models discussed below are given by
\begin{equation}
\sigma_i = \sqrt{F^{-1}_{ii}}\,.
\end{equation}

\section{Forecasted exclusion limits on $r$}\label{sec:rlimits}

\begin{figure}
  \includegraphics[width=3.5in]{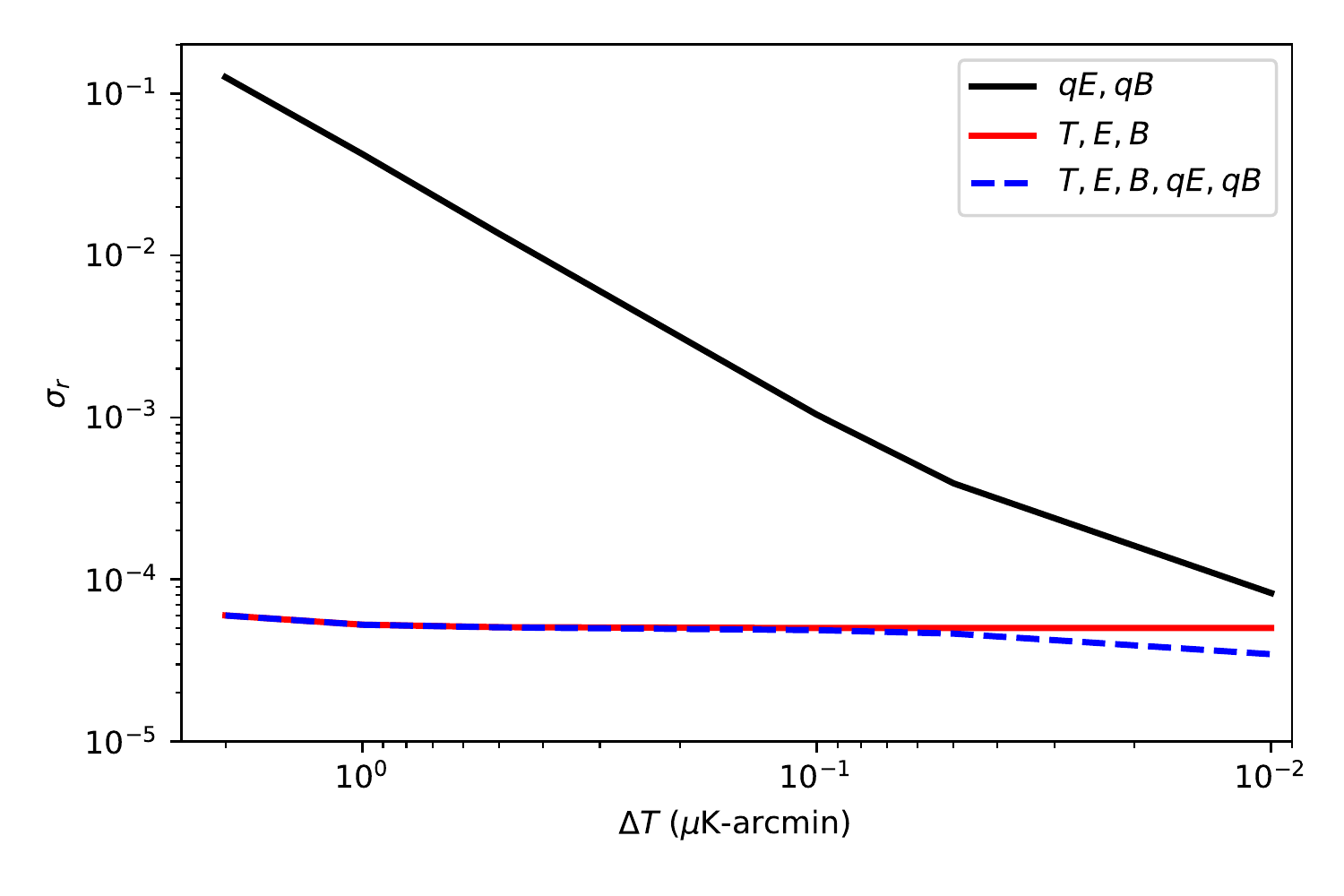}
	\caption{The 1-$\sigma$ constraint on $r$ using $qE, qB$ (black solid), $T, E, B$ (red solid), and $qE, qB, T, E, B$ (blue dashed) as a function of CMB instrumental noise. We assume data on the full sky and an LSST-like galaxy survey. For the remote quadrupole constraint to be competitive with the primary CMB, it is necessary to reach a sensitivity of order $\Delta T \sim 10^{-2} \mu$K-arcmin. }
	\label{fig:exclusion}
\end{figure}

We first forecast exclusion limits on the tensor-to-scalar ratio $r$. In this case, the only parameter in the Fisher matrix is $r$, which takes the fiducial value $r=0$. In principle, the exclusion limit from the remote quadrupole can decrease without bound with increasing sensitivity of the CMB experiment. This is because the reconstruction noise on $qB$, which will dominate the constraint in this limit, goes to zero (see discussion in Sec.~\ref{sec:noise}). However, a strong constraint can already be obtained from the primary CMB $B$ modes. To determine how useful the remote quadrupole field can be in improving the exclusion limit on $r$, 
 we plot in Fig.~\ref{fig:exclusion} the constraint using only the remote quadrupole (black solid), only the primary CMB over the range $2\leq\ell\leq1000$ (red solid), and the joint constraint (blue dashed) as a function of instrumental noise. For the setup considered here, instrumental noise on the order of $10^{-2} \mu$K-arcmin is necessary for the exclusion limit from the remote quadrupole to be competitive with the exclusion limit from the primary CMB. This result is consistent with Ref.~\cite{2012PhRvD..85l3540A}, who obtained similar exclusion limits. Note that the constraint from the primary CMB does not improve significantly with instrumental noise, a consequence of the scalar lensing contributions to $B$. Considering the possibility of de-lensing would both improve the constraint from the primary CMB, but also the constraint from the remote quadrupole, as the reconstruction noise would be lowered (e.g. because $C_{\ell}^{BB}$ in Eq.~(\ref{eq:estimator1noiseq}) would be lowered). We defer a full exploration of this possibility to future work.
 
\section{Constraints on new physics in the primordial tensor sector}
\label{cnf}

Measurements of the remote quadrupole stand to be far more useful for realistic levels of instrumental noise in the case where there is a detectable level of primordial gravitational waves. In this scenario, at low-$\ell$ the sensitivity of CMB experiments will eventually become great enough that measurements of $C_{\ell}^{BB}$ will be cosmic variance or foreground limited due to the limited number of modes. The remote quadrupole fields contain independent information about primordial tensors, and given the limited number of modes at low-$\ell$ in the primary CMB $B$, even a few measured modes of the remote quadrupole field stand to improve our understanding of the tensor sector significantly. To explore this possibility further, we consider two parameterizations for the primordial tensor power spectrum. We first consider the case  where $r$, $n_t$, and $\Delta_c$ are free parameters and the primordial tensor power spectrum behaves according to Eq.~(\ref{eq:rnt}). We then account for a rather different scenario: gravitational waves sourced by gauge field as per the model in \cite{Dimastrogiovanni:2016fuu}, with a maximally chiral ($\Delta_c = 1$)  spectrum:
\begin{equation}\label{eq:sourcedmodel}
\mathcal{P}_h = r_{p} \,\mathcal{P}_\zeta \exp \left[ -\frac{1}{2 \sigma^2} \ln^2 \left( \frac{k}{k_p}\right) \right]\,.
\end{equation}
The tensor-to-scalar ratio at the peak, $r_{p}$, the width of the peak $\sigma$, and the peak scale $k_p$ are free parameters in our forecasts. A value $\Delta_c = 1$ is justified when the total tensor signal around the peak scale is dominated by the sourced contributions, $r=r_{p}$. Note that in general, there are also tensors produced by vacuum fluctuations. Here, we assume that these contributions are subleading.

As mentioned in the introduction, the above parametrization is inspired by the inflationary dynamics in the presence of a (spectator) sector consisting of an axion coupled to gauge-fields. The rolling of the axion produces a transient growth of gauge-field fluctuations which, in turn, leads to the production of chiral gravitational waves~\footnote{Naturally, gauge fields also source scalar curvature fluctuations. The parameter space for this class of models includes large regions where tensor perturbations are sourced for the most part by gauge fields whilst leaving the scalar sector nearly unaffected \cite{Dimastrogiovanni:2016fuu}. {This is the regime being constrained here.}
See for example \cite{Agrawal:2018mrg} for a comprehensive scan of the parameter space for the model in \cite{Dimastrogiovanni:2016fuu}.}. Chirality is a consequence of the parity-breaking gauge-field background. The choice of a Gaussian parametrization for the spectrum mimics the background evolution of the axion: it rolls for a certain number of e-folds sourcing the tensor modes of the gauge fields mostly on scales that happen to exit the horizon during that time.
This type of dynamics is realized in several A-G inflationary set-ups. In particular, the parametrization of Eq.~(\ref{eq:sourcedmodel}) fits the predictions of the model put forward in \cite{Dimastrogiovanni:2016fuu} rather well, where the gauge field is an SU(2) multiplet and the Lagrangian reads:
\begin{equation}
\mathcal{S}=\int d^4x\sqrt{-g}\left[\frac{M_{\text{Pl}}^{2}}{2}R+\mathcal{L}_{\phi}-\frac{1}{2}\left(\partial\chi\right)^{2}-U(\chi)-\frac{1}{4}F_{\mu\nu}^{a}F^{a\mu\nu}+\frac{\lambda\,\chi}{4f}F_{\mu\nu}^{a}\tilde{F}^{a\mu\nu}\right]\,,
\end{equation}
where $\mathcal{L}_{\phi}$ stands for the Lagrangian of the inflaton field, $\chi$ is the axion, $F_{\mu\nu}^{a}\equiv\partial_{\mu}A^{a}_{\nu}-\partial_{\nu}A^{a}_{\mu}-g \epsilon^{abc} A^{b}_{\mu}A_{\nu}^{c}$ and $\tilde{F}^{a\mu\nu}\equiv\epsilon^{\mu\nu\rho\sigma}F^{a}_{\rho\sigma}/(2\sqrt{-g})$.\\ It's important to stress at this stage that the parameterization in Eq.~(\ref{eq:sourcedmodel}) is not restricted to one set-up and captures rather well also  the U(1) case (see e.g. \cite{Barnaby:2012xt,Mukohyama:2014gba,Namba:2015gja,Peloso:2016gqs}). As expected, an Abelian \textit{vs} non-Abelian choice results in distinct predictions and often in a richer phenomenology for the non-Abelian case. However, these differences are not conspicuous at the level of the tensor power spectrum.

\subsection{Chiral tensor model with $r$, $n_t$, and $\Delta_c$}

\begin{figure}
  \includegraphics[width=7in]{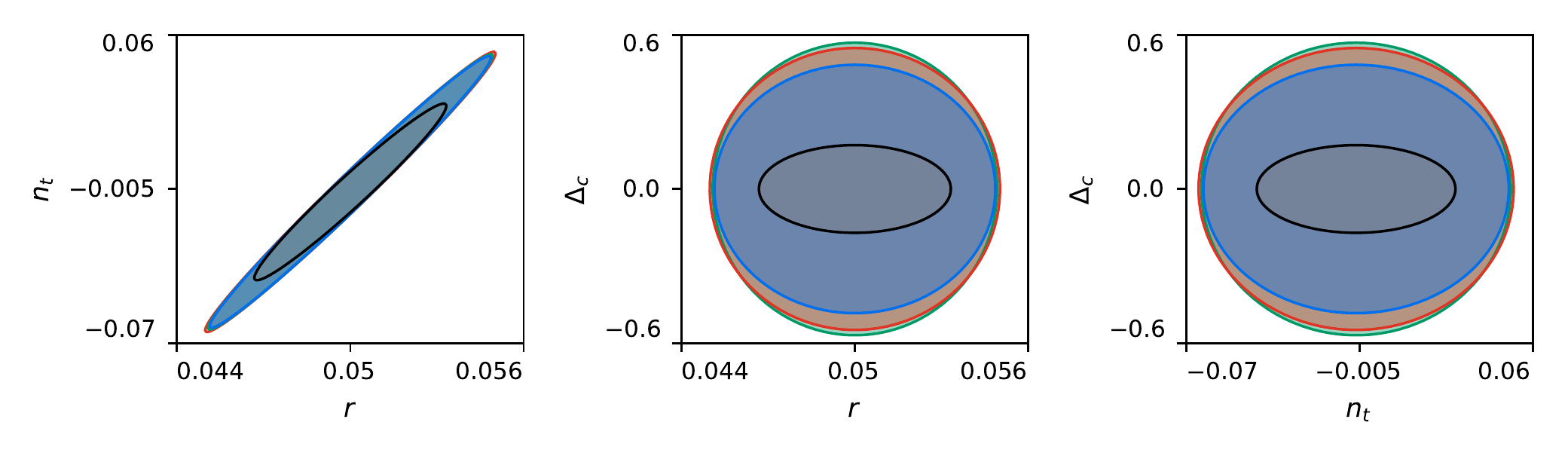}

	\caption{Comparison of the $1-\sigma$ constraints on the three-parameter tensor sector for noiseless measurements of the primary CMB for $2 < \ell < 1000$ (Green ellipse) and the joint constraint including the remote quadrupole for CMB experiments with sensitivity $1 \mu$K-arcmin (Red ellipse), $0.1 \mu$K-arcmin (Blue ellipse), and no CMB noise (Grey ellipse). We assume central values $r=0.05$, $\Delta_c = 0$, $n_t = -r/8$.}
	\label{fig:highell_comparison}
\end{figure}

For our first model, we assume the power-law spectrum Eq.~(\ref{eq:rnt}) with $r$, $n_t$, and $\Delta_c$ as free parameters taking the central values $r=0.05$, $n_t=-r/8$, and $\Delta_c=0$. The underlying model is non-chiral with a tensor tilt given by the single-field inflationary consistency condition, and we wish to forecast how well one can measure $r$ and $n_t$ and how well one can constrain $\Delta_c$. In Fig.~\ref{fig:highell_comparison}, we show the forecasted constraints for this fiducial model using only the primary CMB $T,E,B$ in the zero-noise cosmic variance limit (Green), compared to the forecast including $T,E,B, qE, qB$ for varying levels of instrumental noise: $\Delta T = 1 \mu$K-arcmin (Red), $\Delta T = 0.1 \mu$K-arcmin (Blue), and $\Delta T = 0 \mu$K-arcmin (Grey). As can be seen in this plot, it is possible to use the additional information in the remote quadrupole fields to beat the cosmic variance limit from the primary CMB alone. The greatest improvement is for the chirality parameter $\Delta_c$, which is consistent with previous work on constraining chirality with the primary CMB. As shown in Ref.~\cite{Gluscevic:2010vv}, there is little constraining power on chirality in the primary CMB for $\ell > 10$, making constraints limited by the number of modes. This is precisely the scenario where the additional information from the remote quadrupole fields on large scales is most useful. There is also modest improvement on $r$ and $n_t$, again from the additional independent modes. 

Next, we consider the scenario of a maximally chiral model with $r=0.05$, $n_t=-r/8$, and $\Delta_c=1.0$. In this case, using the primary CMB alone, for $\Delta T = 0.01 \ \mu$K-arcmin, the marginalized constraint on $\Delta_c$ is $\sigma_{\Delta_c} \simeq 0.7$. Adding the remote quadrupole fields for the same level of CMB instrumental noise, this shrinks to $\sigma_{\Delta_c} \simeq 0.53$. This is a non-trivial improvement, however, it would only take a $\sim 1.5\sigma$ detection of chirality to a $\sim 2.0\sigma$ detection -- this would be far from conclusive evidence. In the absence of dramatically lower instrumental noise, this suggests that other techniques, such as those proposed in Ref.~\cite{Masui:2017fzw}, would be necessary to definitively establish that PGWs are chiral at this amplitude.

For models with a significant blue tilt ($n_t \sim \mathcal{O}(0.1)$), there is very little contribution from the remote quadrupole to the constraints on $r$ and $n_t$, even in the cosmic variance limit. In addition, the overall constraint on $\Delta_c$, and the improvement possible with the remote quadrupole, weaken significantly. In both cases, this is because the remote quadrupole field receives contributions from relatively low-$k$, and adding a blue tilt significantly lowers the amplitude of the quadrupole field, rendering the tensor contribution even more difficult to detect. We therefore conclude that pSZ tomography does not provide extra constraining power for models with a blue tilt.

\subsection{Sourced chiral tensor model}

\begin{figure}
  \includegraphics[width=6in]{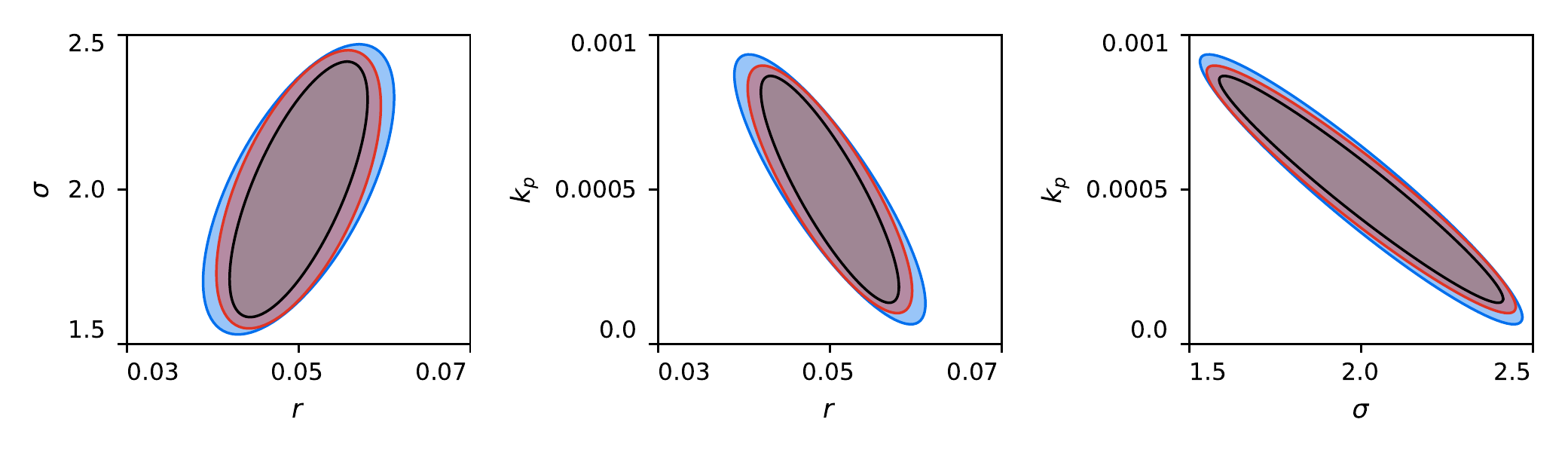}
  \includegraphics[width=6in]{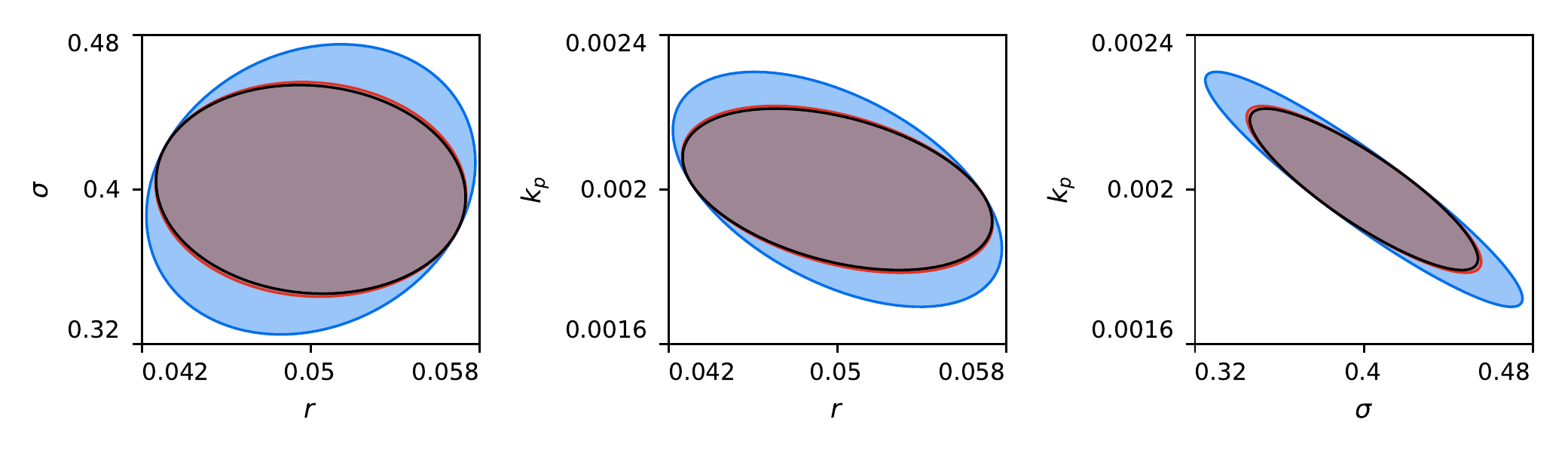}

	\caption{Constraints on the sourced chiral tensor model. The blue ellipse is the constraint including the low-$\ell$ and high-$\ell$ CMB $T,E,B$ for an experiment with $\Delta T = 1 \mu$K-arcmin. The red ellipse shows the constraints including $qE, qB$ also for $\Delta T = 1 \mu$K-arcmin. The grey ellipse is the cosmic variance limited constraint including the low-$\ell$ and high-$\ell$ CMB $T,E,B$ and $qE, qB$.}
	\label{fig:bump_model}
\end{figure}

We now consider the sourced chiral model with power spectrum Eq.~(\ref{eq:sourcedmodel}), representing A-G inflationary models. In Fig.~\ref{fig:bump_model}, we show the constraints on the three free parameters $r, \sigma, k_p$ for two fiducial parameter sets: $\{0.05, 2, 0.0005 \}$ and $\{0.05, 0.4, 0.002 \}$. The first set is for a broad bump on large scales, while the second is for a narrow bump on somewhat smaller scales. The constraint from the CMB only for $\Delta T = 1 \mu$K-arcmin is shown in blue. The constraint including the remote quadrupole fields for the same noise level is shown in red, and the cosmic variance limited constraint in grey. In both cases, there is some improvement on the constraints when including the remote quadrupole fields. Furthermore, this improvement saturates at relatively modest sensitivity. More generally, we find that constraints are affected the most by the addition of the remote quadrupole fields for smaller values of $\sigma$ and values of $k_p$ in the range considered here. This is due to the limited range in scales probed by the measurable (low-$\ell$) moments of the remote quadrupole field. Should such models be realized in nature, pSZ tomography could therefore be a useful tool for improving constraints on A-G inflationary models.

\section{Altered tensor growth function}\label{sec:alteredgrowth}

In principle, pSZ tomography opens a new observational window to the decay of tensor modes as they re-enter the horizon. This is because the qE and qB fields are reconstructed in a set of tomographic redshift bins, implying that the decay of tensors can be probed by comparing the amplitude of these fields among bins. Above, we have parameterized the evolution of tensors through the tensor transfer function $D^T (k, \chi)$, assuming decay consistent with a matter dominated Universe. In $\Lambda$CDM there will be additional decay from the cosmological constant beyond what we have considered here. Additionally, in a number of modifications of general relativity, the decay of tensors at late times can be altered. A dramatic example is massive bigravity~\cite{Cusin:2014psa,Fasiello:2015csa,Johnson:2015tfa}, where decay can even turn around and lead to tensors growing in time! In more vanilla models of dark energy, where the equation of state and its time-variation are considered (e.g. quintessence), the decay of tensors can be altered. In order to explore the ability of pSZ tomography to constrain the tensor transfer function, we employ a one-parameter phenomenological model where the late-time decay of tensors on all scales is enhanced:
\begin{equation}\label{eq:transferphi}
 D^T (k,\chi) = 3 \frac{j_1 (k \chi)}{k\chi} \exp \left[ -\phi \left( \frac{\tau}{\tau_0} \right) \right],
\end{equation}
Here, $\phi$ is a dimensionless constant and $\tau_0$ is the present conformal time. We do not expect this phenomenological model to match any of the previously mentioned scenarios very well, but it should give a reasonable idea of the sensitivity of pSZ tomography to modifications of the tensor transfer function at late times.

\begin{figure}
  \includegraphics[width=6in]{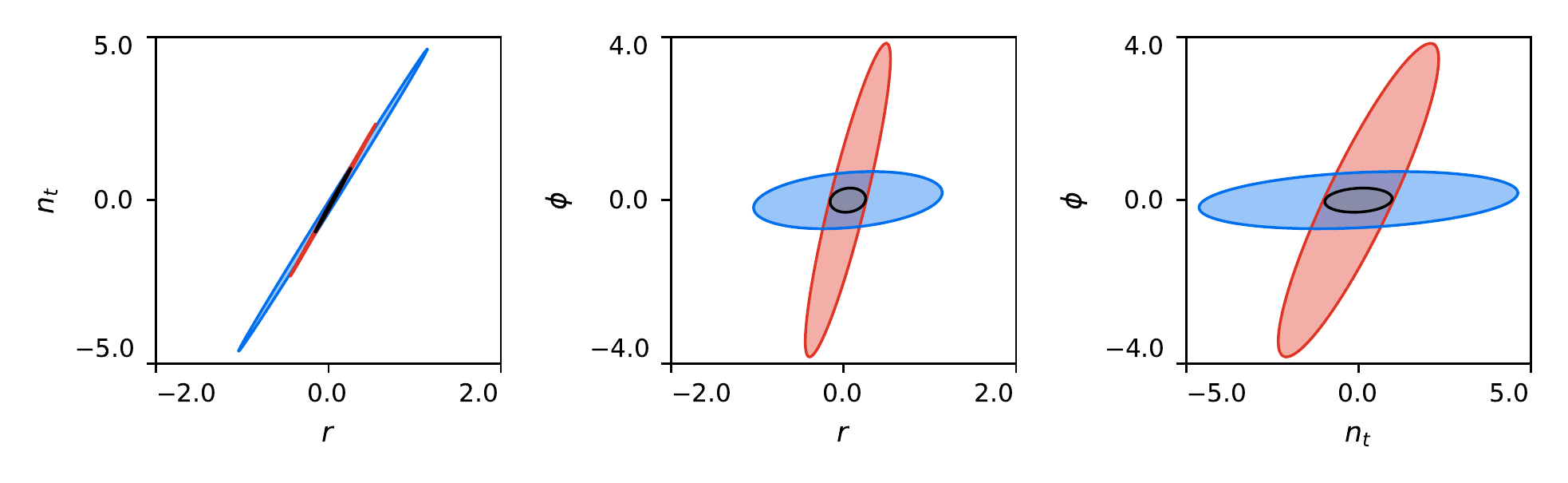}
	\caption{Marginalized constraints on $r$, $n_t$, and a parameter $\phi$ modifying the tensor transfer function as in Eq.~\ref{eq:transferphi}. Constraints from the primary CMB for $2\leq \ell \leq 10$ are in red, pSZ in 12 redshift bins in blue, and the joint constraints in grey. Throughout we assume an instrumental noise of $\Delta T=0.1 \ \mu$K-armin.}
	\label{fig:tensorgrowth}
\end{figure}

In Fig.~\ref{fig:tensorgrowth} we show the marginalized constraints on $r$, $n_t$, and $\phi$ using the CMB (red), qEqB (blue), and the joint constraints (grey) using multipoles $2 < \ell < 10$, 12 redshift bins, and an instrumental noise $\Delta T = 0.1 \ \mu$K-arcmin. Note that since it is the late-time evolution of tensors that is affected, there is little additional constraining power from the higher multipoles of the CMB, which are sourced primarily at recombination. As expected, since the primary CMB arises from a line of sight integral, there are significant degeneracies between the primordial power spectrum and the late-time decay of tensors. The redshift information in the remote quadrupole breaks this degeneracy. This makes pSZ tomography an excellent probe of the late-time evolution of tensors. Exploring instrumental noise levels $\Delta T = \{ 1, 0.1, 0.01\} \ \mu$K-arcmin we obtain joint marginalized constraints on $\phi$ using both the CMB and remote quadrupole given by $\sigma_\phi = \{ 1.4, 0.29, 0.065 \}$ as compared to the CMB-only constraint of $\sigma_\phi  = 3.8$. For ambitious levels of instrumental noise, pSZ tomography can probe the tensor transfer function at the percent level. Furthermore, the constraints improve monotonically with decreasing CMB noise, implying that if we do indeed detect tensors, the remote quadrupole field from tensors could eventually provide a probe of modified gravity and dark energy much in the same way that measurements of the scalar growth function does.

\section{Conclusions}
\label{conclusions}

In this paper we have assessed the impact that pSZ tomography may have on understanding new physics in the tensor sector. The quadrupole fields reconstructed using pSZ tomography carry new information beyond the primary CMB temperature and polarization anisotropies. Furthermore, the B-mode remote quadrupole field is in principle a ``clean'' probe of tensors, since unlike the primary CMB B-modes, it does not receive contributions from lensing. However, the experimental requirements to obtain a high fidelity reconstruction of the remote quadrupole fields, even in the idealized case considered here of full-sky data and in the absence of foregrounds and systematics, are rather futuristic. CMB instrumental noise levels of order $\Delta T \sim 10^{-2} \ \mu$K-arcmin are necessary for pSZ tomography to yield comparable exclusion bounds to those in principle obtained by the primary CMB in the absence of delensing ($\sigma_r \sim 5 \times 10^{-5}$). A similar level of instrumental noise is necessary to significantly improve the constraints on chirality of primordial gravitational waves, although somewhat less stringent requirements are necessary to improve constraints on Axion Gauge field inflation models (around $\Delta T \sim 1 \ \mu$K-arcmin). However, should tensors be detected, ambitious but perhaps achievable levels of instrumental noise (around $\Delta T \sim 0.1-1 \ \mu$K-arcmin) are necessary to probe the late-time decay of tensors at the $\sim 10\%$ level.

Because our focus has been on assessing how informative pSZ tomography could be in principle, we have neglected a number of important real-world effects that could significantly affect what is achievable in practice. First, we have assumed that we have access to data on the full sky. Accounting for cuts associated with galactic and extra-galactic foregrounds will degrade the measurement of the low-$\ell$ CMB as well as the reconstructed remote quadrupole fields. This can be particularly impactful for the remote quadrupole since most of the signal is at $\ell=2$. For example, Ref.~\cite{2012PhRvD..85l3540A} found that with $f_{\rm sky}=0.83$ the total $(S/N)^2$ of the remote quadrupole field is reduced by a factor of $\sim 3$ (as opposed to $1/f_{\rm sky}$ as would naively been expected); larger sky cuts yield larger reductions. While sky cuts could be mitigated by better modelling of foregrounds, this is clearly a challenge. We have also neglected foreground residuals, both in our assessment of the detectability the primary CMB and in the reconstruction noise of the remote quadrupole. Galactic foregrounds in polarization on large angular scales have proven to be particularly insidious, affecting inferences of the low-$\ell$ polarization power spectra (as evidenced by the significant analysis required to understand the low-$\ell$ polarization results from Planck). Because pSZ tomography lies on the sensitivity and resolution frontier, which is only now being accessed, significantly less is understood about foregrounds on the very small angular scales ($\ell \sim \mathcal{O} (10^3 -10^4)$) relevant for the reconstruction noise. On the optimistic side, one might hope that the foreground problem at high-$\ell$ could be more readily understood and modelled than the foreground at low $\ell$. Another important systematic is the so-called optical depth degeneracy, a multiplicative bias on the reconstructed quadrupole field due to imperfect knowledge of the galaxy-electron cross power, $C_{\alpha, \ell}^{\Delta \tau \delta_g}$. This introduces a nuisance parameter which must be marginalized over in each redshift bin, which will have the effect of weakening the constraints presented here (in particular the constraints on modified late-time decay of tensors).

We have also neglected other obvious systematics, including redshift errors in the galaxy survey (not expected to be important in the broad bins we consider here), anisotropic noise in the CMB experiment, biases in the quadratic estimator used to reconstruct the remote quadrupole fields, etc. While many of these effects could be mitigated in principle, for example through detailed modelling, a dedicated analysis is necessary to determine what the limiting factor in using pSZ tomography to constrain the tensor sector will be.  

Although the experimental requirements are stringent, and overcoming possible systematics a daunting task, we stress that information about tensor modes on horizon scales is limited and difficult to measure by other means. For example, tensors can be measured through their tidal effects on density perturbations on very large scales~\cite{2010PhRvL.105p1302M,Dai:2013kra,Schmidt:2013gwa}, through the lensing of 21-cm fluctuations~\cite{Book:2011dz}, or via cosmic shear in weak lensing surveys~\cite{2010PhRvD..82b3522D,Chisari:2014xia}. The former two methods require futuristic 21cm dark ages surveys, while the latter is only plausible for observationally ruled out values of $r$, in the absence of large scale dependence. pSZ tomography, on the other hand, could in principle yield some new information on tensors with planned next-generation galaxy surveys and CMB experiments, should we be just on the cusp of detecting $r$. In addition, even a low-significance detection of gravitational waves using pSZ tomography would provide an important independent check on a detection found from the primary CMB B modes as it would involve different foregrounds and systematics. Given that the necessary datasets will become available in the coming decades, the approach of pSZ tomography should be kept in mind as a viable alternative probe of the tensor sector that should be developed further.

\acknowledgments
\noindent AD is supported by NSF Award PHY-1417385. ED is supported in part by DOE grant DE-SC0009946. MRF acknowledges support from STFC grant ST/N000668/1. ED and MRF are delighted to thank the Perimeter Institute for Theoretical Physics for hospitality and support whilst this work was in progress. MCJ is supported by the National Science and Engineering Research Council through a Discovery grant. This research was supported in part by Perimeter Institute for Theoretical Physics. Research at Perimeter Institute is supported by the Government of Canada through the Department of Innovation, Science and Economic Development Canada and by the Province of Ontario through the Ministry of Research, Innovation and Science. We acknowledge the use of the CosmicFish code~\cite{Raveri:2016leq,Raveri:2016xof} in producing some of the results in this paper.

\bibliography{psztensor}

\end{document}